\documentclass[fleqn,twoside]{article}%
\usepackage{amsmath}
\usepackage{amsfonts}
\usepackage{amssymb}
\usepackage{graphicx}%
\def\be{\begin{equation}}

\def\ee{\end{equation}}
\def\bes{\begin{equation}\begin{split}&}
\def\es{\end{split}}
\def\bi{\bibitem}

\begin{document}
\title{\boldmath Why scalar-tensor equivalent theories are not physically equivalent?}
\author{Nayem Sk$^\dag$, Abhik Kumar Sanyal$^\ddag$}
\maketitle
\noindent
\begin{center}
\noindent
$^{\dag}$ Dept. of Physics, University of Kalyani, West Bengal, India - 741235.\\
\noindent
$^{\ddag}$ Dept. of Physics, Jangipur College, Murshidabad,
\noindent
West Bengal, India - 742213. \\

\end{center}
\footnotetext[1] {\noindent
Electronic address:\\
\noindent
$^{\dag}$nayemsk1981@gmail.com\\
$^{\ddag}$sanyal\_ ak@yahoo.com}

\begin{abstract}
Whether Jordan's and Einstein's frame descriptions of F(R) theory of gravity are physically equivalent, is a long standing debate. However, practically none questioned on true mathematical equivalence, since classical field equations may be translated from one frame to the other following a transformation relation. Here we show that, neither Noether symmetries, Noether equations, nor may quantum equations be translated from one to the other. The reason being, - conformal transformation results in a completely different system, with a different Lagrangian. Field equations match only due to the presence of diffeomorphic invariance. Unless a symmetry generator is found which involves Hamiltonian constraint equation, mathematical equivalence between the two frames appears to be vulnerable. In any case, in quantum domain Mathematical and therefore physical equivalence can't be established.

\end{abstract}

\maketitle
\flushbottom

\section{Introduction}
Modified theory of gravity, in particular its $F(R)$ counterpart, is a strong contender to an alternative theory of dark energy, required for late-time accelerated expansion of the universe \cite{a, b}. The action for such a theory is expressed as

\be\label{1a} A = \int\sqrt{-g} d^4 x \left[{F(R)\over 2\kappa} + \mathcal{L}_m\right] +\Sigma\ee
where, $\mathcal{L}_m$ is the matter lagrangian, $\kappa = 8\pi G$, and $\Sigma = {1\over \kappa}\int_{\partial\mathcal{V}}\sqrt{h} ~d^3 x F'(R) K$ is the supplementary boundary term, while prime denotes derivative with respect to the Ricci Scalar $R$. It's customary to invoke its scalar-tensor equivalent form, and constrain the mass of a fictitious field - the scalaron, so that a particular $F(R)$ model passes the solar test and admits cosmological bound simultaneously \cite{a} - \cite{Bar}. Canonical formulation of the theory to scalar tensor equivalent form, either in Jordan's frame of reference or in Einstein's frame of reference is also required to find associated symmetries and for canonical quantization. Translation to the Jordan's frame of reference is possible, simply under redefinition of $F'(R) = \Phi$ and $R = U_{,\Phi}$. On the other hand, translation of the action \eqref{1a} in Einstein's frame of reference is possible, under conformal transformation $\tilde{g}_{\mu\nu} = F'(R) g_{\mu\nu} = e^{2\omega} g_{\mu\nu}$, where the conformal factor $\omega$ is related to an effective scalar field $\tilde\phi$ by the relation $\omega = \sqrt{k\over 6}\tilde\phi.$ Both the techniques yield correct classical field equations, and mathematical equivalence is established through the transformation,

\be\label{2rel}\tilde \phi = \sqrt{3\over 2k} \ln \Phi.\ee
It is therefore tacitly assumed that, not only scalar tensor equivalent forms are identical with each other, but also, they represent the same dynamics of the action (\ref{1a}). Still there is a controversy. Firstly, unlike scalar-tensor equivalence in Einstein's frame, matter sector in Jordan's frame remains unaltered under the said transformation, i.e., matter action $S_m$ remains independent of the scalar field $\Phi$. Therefore the weak equivalence principle holds, and test particles follow geodesics lines of $g_{\mu\nu}$ in Jordan's frame. This signals that these two techniques might not be physically equivalent. In fact physical equivalence of the two has been questioned over decades \cite{2} - \cite{SSV}, while the mathematical equivalence has been trivially accepted, since classical field equations may be translated back and forth, using the transformation relation \eqref{2rel}. However, true mathematical equivalence is established, only if both the frames yield the same mathematical results independent of each other, or at least if one can switch the results back and forth between the two frames. Let us be more precise to understand what sort of mathematical results we are talking of. The field equation corresponding to action \eqref{1a} reads

\be\label{3f}F'(R)R_{\mu\nu}- \frac{1}{2}F(R)g_{\mu\nu} - \nabla_\mu \nabla_\nu F'(R) + g_{\mu\nu}\Box F'(R) = \kappa T_{\mu\nu}\ee
where, $\Box F'(R)= {F'~^{;\mu}}_{;\mu} = {1\over \sqrt{-g}}\partial_{\mu}(\sqrt{-g}g^{\mu\nu}\partial_{\nu}F')$.
The trace of the above equation (\ref{3f})
\be\label{3t1}
RF'-2F+3F'^{~;\mu}_{~~~~;\mu}=\kappa T
\ee
may be expressed in the context of homogeneous space-time as,
\be\label{3t2}
RF'-2F+3\left[(\ddot R+\theta \dot R)F''+\dot R^2F'''\right]=\kappa T^\mu_{\mu} = \kappa T
\ee
where, $\theta = v^{\alpha}_{;\alpha} = {1\over \sqrt{-g}}{d\over dt}(\sqrt{-g})$ is the expansion scalar, and $v^{\alpha}$ are the components of four velocity vector. Now under the choice $F(R)\propto R^n$, and for trace-less matter field (pure vacuum or radiation era), the trace equation (\ref{3t2}) may be expressed as
\bes\label{3t3}
(n-2)R^n+3(\ddot R+\theta \dot R)n(n-1)R^{n-2}
 + 3\dot R^2 n(n-1)(n-2)R^{n-3}=0.
\end{split}\ee
The above equation is satisfied for $n = 2$, i.e.
\bes\label{3t4}
F = F_0 R^2,~ \mathrm{if},~(\ddot R+\theta \dot R)=0,~
\mathrm{i.e.},~ \mathfrak{F} = \dot R\sqrt{-g}
\end{split}\ee
Hence, action \eqref{1a} admits a general symmetry ($\mathfrak{F}$ = constant, independent of metric), and the reason is that, $F = F_0 R^2$ leads to a scale invariant action \cite{schmidt}. This is the result we are talking of. The above symmetry is also expected to emerge from the canonical actions corresponding to \eqref{1a} (formulated either following Lagrange multiplier technique, or through scalar-tensor equivalence in Jordan's frame and also in Einstein's frame), once Noether analysis is performed. Further, Noether analysis might yield additional symmetries, which must also be realized from different canonical actions. However, Noether symmetry analysis performed to find a form of the potential of a theory, initiated by de. Ritis et al \cite{rug}, or to find a suitable form of $F(R)$, as in the present case, is applicable only with finite degrees of freedom. So it is required to take up a particular metric, and for the present purpose, let us consider isotropic and homogeneous Robertson-Walker metric,

\be\label{rw} ds^2 = -dt^2 + a(t)^2 \left[{dr^2\over 1-kr^2} + r^2 d\theta^2 + r^2 \sin^2 \theta d\phi^2\right].\ee
Now, as already mentioned, true mathematical equivalence is established only if both the frames admit identical symmetries independently, or if at least such symmetries may be translated from one frame to the other, following the transformation relation (\ref{2rel}). In this respect, the aim of the present work is to show that, the frames are not truly mathematically equivalent, since the point Lagrangians obtained following Lagrange multiplier technique and also translating to the Jordan's don't admit the Noether counterpart of the symmetry \eqref{3t4}. Further, the two frames yield different additional symmetries, which can't be switched back and forth. This, we establish in the following section. In section 3, we shall also take up canonical quantization scheme and translate the quantum dynamical equation from one frame to the other, to establish the fact that the frames are not even dynamically equivalent in the quantum domain. To be more precise, the two Hamiltonian are not related under canonical transformation. In section 4, we expatiate the underlying reason for such in-equivalence. We conclude in section 5.\\

\section{Noether symmetries}

Associated Noether symmetry corresponding to action \eqref{1a} to find a form of $F(R)$, has been addressed over decades \cite{lmn1} - \cite{lmn8}. However, the canonical action required to address such symmetry, was formulated following Lagrange multiplier technique, treating $R - 6\left({\ddot a\over a}+{\dot a\over a}^2 + {k\over a^2}\right)$, as a constraint of the theory, in Robertson-Walker metric \eqref{rw}. The point Lagrangian thus obtained reads

\be \label{cl}\mathcal{L} = {1\over 2\kappa} \left[a^3(F-R F') -6a(\dot a^2-k) F' -6a^2 \dot a\dot R F''\right] + \mathcal{L}_m.\ee
The only result obtained invoking Noether symmetry, is $F(R) \propto R^{3\over 2}$. Even enlarging the configuration space by introducing a scalar field sector in action \eqref{1a}, and/or working in anisotropic models yield no new result, other than $R = $ constant \cite{lmn9}. Firstly, there is absolutely no chance to retrieve the general symmetry \eqref{3t4}, since this technique of canonical formulation is applicable with finite degrees of freedom, and therefore, not a general formalism. Next, even the Noether counterpart of the general symmetry \eqref{3t4}, i.e. $\mathfrak{F} =  a^3 \dot R$, for $F(R) \propto R^2$ also remains absent. It is therefore clear that, the above canonical form \eqref{cl}, neither gives the complete mathematical nor the dynamical picture of the system \eqref{1a} under consideration. However, possibility of obtaining additional symmetries might emerge following canonical formulation of the action \eqref{1a}, in scalar-tensor equivalent forms, which we shall consider in the following subsections.

\subsection{Symmetries in Jordan's Frame}
\noindent
Scalar-tensor equivalent form in Jordan's frame of reference is established in \cite{a} - \cite{Sot}, and also in \cite{12, 13}, by first expressing the action \eqref{1a} in the following form.

\be\label{j1} S = \frac{1}{2\kappa} \int  d^4 x\sqrt{-g}[ F(\chi)+ F'(\chi)(R-\chi)]+ S_m + \Sigma,\ee
where prime denotes differentiation with respect to $\chi = R$. Now variation with respect to $\chi$, leads to the relation $F''(R)(\chi-R)=0$, and for $F''(R)\ne0$, the choice of the auxiliary variable $\chi=R$, is regained. Finally, redefining the field $\Phi$ and its potential $U(\Phi)$ by

\be\label{j2} \Phi =F'(R);~~~~ U(\Phi)= \chi(\Phi)\Phi-F(\chi(\Phi))\ee
the action (\ref{j1}) takes the form

\be\label{j3} S_{Jor} = \frac{1}{2\kappa} \int  d^4 x\sqrt{-g}[ \Phi R-U(\Phi)]+S_m + \Sigma.\ee
Action (\ref{j3}) therefore establishes dynamical equivalence between $F(R)$ theory and a class of Brans-Dicke theories, with Brans-Dicke parameter $\omega = 0$, and a potential $U(\Phi)$. Under metric variation, the counter term $\Sigma$ gets cancelled with the boundary term, and the field equations read,

\begin{subequations}\begin{align}
&\label{j4}
\Phi\left(R_{\mu\nu}-\frac{1}{2}g_{\mu\nu}R\right)+{\Phi^{;\alpha}}_{;\alpha}g_{\mu\nu}-\Phi_{;\mu;\nu}+
\frac{1}{2}g_{\mu\nu}U(\Phi)=\kappa T_{\mu\nu}\\
&\label{j5}
U_{,\Phi}(\Phi) = R
\end{align}\end{subequations}
It is important to mention that the trace of the field equations admits the general conserved current $\mathfrak{F} = \dot R\sqrt{-g}$, for $F(R) \propto R^2$ presented in equation \eqref{3t4}. Now, in the Robertson-Walker metric \eqref{rw} the point Lagrangian reads
\bes\label{j6}
L_t(a,\Phi,\dot a,\dot \Phi) =\frac{1}{16\pi G} [- 6a^2 \dot a \dot \Phi - 6a \dot a^2\Phi+ 6\Phi a k- a^3U(\Phi)]
 -\rho_0 a^{-3w},
\end{split}\ee
where we have taken barotropic fluid inclusive of dark matter, $w$ being the state parameter, and $\rho_0$ is a constant representing the present value of energy density content of the universe. In the Noether symmetry approach, the lift vector $X$ acts as an infinitesimal generator of Noether symmetry in the tangent space $(a,\dot a,\Phi,\dot \Phi)$,
\be\label{j7} X = \alpha \frac{\partial }{\partial a}+\beta\frac{\partial }{\partial \Phi}
+\dot\alpha \frac{\partial }{\partial\dot a}+ \dot\beta\frac{\partial }{\partial\dot \Phi}, \ee
The existence condition for symmetry, $\pounds_X L =X L = 0 $, then leads to the following system of partial differential equations,
\bes\label{j8}
   \alpha_{,\Phi} = 0; \;\;\; \Phi\alpha+a\beta +2a\Phi\alpha_{,a}+a^2\beta_{,a} = 0; \;\;\;2\alpha + a \beta_{,\Phi}+a\alpha_{,a}= 0; \\&
   3\alpha \left({2k\Phi-a^2 U\over 2\kappa} - w\rho_0 a^{-(3w+1)}\right)+{a\beta\over 2\kappa}(6k-a^2 U_{,\Phi})  = 0.
\end{split}\end{equation}
The symmetries admissible in this frame (see appendix) are found to be the same as obtained in view of following Lagrange multiplier technique. Both in vacuum ($\rho = 0 = p$) and in pressure-less dust ($p = 0$), for $k = 0, \pm 1$, as, $F(R) = F_0 R^{3\over 2}$, which carries a conserved current $\mathfrak{F} = {d\over dt}(a\Phi)={d\over dt}(a\sqrt R)$. However, the Noether counterpart of the general symmetry viz. $F = F_0 R^2$ for $\mathfrak{F} = a^3 \dot R$ in vacuum and radiation era remains absent. Thus, in Jordan's frame of reference the general symmetry \eqref{3t4} has been retrieved. However, no Noether symmetry other than the one obtained following Lagrange multiplier technique is found. Thus the two frames appear to be equivalent.\\

\subsection{Symmetries in Einsteins's Frame}
Let us now turn our attention to scalar-tensor equivalent form of action (\ref{1a}) in Einstein's frame. Under the conformal transformation \cite{a} -\cite{Sot}, \cite{7}, and \cite{13} - \cite{17}
\be\label{e1}\tilde{g}_{\mu\nu}= e^{2\omega} g_{\mu\nu}= {\Omega}^2  g_{\mu\nu},\;\;\;
\mathrm{where},\;\;\; \Omega^2=F'(R)= e^{\sqrt{2\kappa\over 3}\tilde \phi},\ee
the action \eqref{1a} may be cast as
\bes\label{e2} A = \int d^4 x\sqrt{-\tilde g}\Big[\frac{1}{2\kappa}\tilde R - \frac{1}{2}\tilde\phi_{;\alpha}\tilde\phi^{;\alpha}-\tilde V(\tilde\phi)\Big]
+\int d^4x {\mathcal{\tilde L}}_m\Big({{\tilde g}_{\mu\nu}\over F'(\tilde\phi)}, \psi_m\Big),\end{split}\ee
where, $\psi_m$ stands for additional matter field in any of its form, and the potential is
\be\label{e3} \tilde V(\tilde\phi) = \frac {R F'(R)-F(R)}{2\kappa(F'(R))^2}.\ee
The field equations are
\begin{subequations}\begin{align}
&\label{fe1} \tilde G_{\mu\nu} =  8\pi G \big(\tilde T_{\mu\nu}^{(m)}+\tilde T_{\mu\nu}^{(\tilde\phi)}\big),\\
&\label{fe2}\tilde{\Box} \tilde\phi - \tilde V_{,\tilde\phi} -\sqrt{4\pi\over 3}\tilde T^{(m)} = 0,\end{align}\end{subequations}
where,

\bes\label{em1} \tilde T^{(m)}_{\mu\nu} = {2\over \sqrt{-\tilde g}} {\delta\over\delta\tilde g^{\mu\nu}} \left(\sqrt{-\tilde g}\tilde{\mathcal{L}}_m\right)=(\tilde\rho + \tilde p)\tilde u_\mu\tilde u_\nu + \tilde p\tilde g_{\mu\nu},\\&
\tilde T_{\mu\nu}^{(\tilde\phi)} =
\tilde\phi_{,\mu}\tilde\phi_{,\nu}-\tilde g_{\mu\nu}\Big[{1\over 2}\tilde g^{\alpha\beta}\tilde\phi_{,\alpha}\tilde\phi_{,\beta}+\tilde V(\tilde\phi)\Big],\end{split}\ee
with $\tilde u^\mu = \Omega^{-1} u^\mu$. So unlike Jordan's frame, scalar field is now coupled to matter. For trace-less matter source, the gravitational action with a constant potential $\tilde V(\tilde\phi)=V_0$, yields $F = F_0 R^2$ in view of \eqref{e3}. Correspondingly, $\tilde{\Box}\tilde\phi = {1\over \sqrt{-\tilde g}}\partial_{\mu}\big(\sqrt{-\tilde g}\tilde g^{\mu\nu}\tilde\phi_{,\nu}\big) =0$, in view of scalar field equation \eqref{fe2}. This leads to a conserved current in homogeneous cosmology in the form, $\mathfrak{F} = \sqrt{-\tilde g}\tilde g^{00}\tilde\phi_{,0}$, which may be translated in proper time to, $\mathfrak{F} = \dot R\sqrt {-g}$, using the transformation relation \eqref{e1}. Thus, Einstein's frame also admits the general conserved current \eqref{3t4}. The point Lagrangian corresponding to the above action \eqref{e2} in RW metric \eqref{rw} reads

\bes\label{e4} \tilde L_{\tilde t}[\tilde a,\tilde \phi,\dot{\tilde a},\dot{\tilde \phi}]=\frac{3}{\kappa}(-\tilde a \dot{\tilde a}^2
+k\tilde a)+\left({1\over 2}\dot{\tilde\phi}^2 - \tilde V(\tilde\phi)\right)\tilde a^3
- \rho_0\tilde a^{-3w}e^{-{\sqrt {\kappa\over 6}}(1-3w)\tilde \phi}.\end{split}\ee
In the above, dot represents, derivative with respect to $\tilde t$. Now, introducing the Noether vector field $\bar X_{\tilde t}$ relative to the Lagrangian $\tilde L_{\tilde t}$ (\ref{e4}) in the tangent space $[\tilde a,\dot{\tilde a},\tilde\phi,\dot{\tilde \phi}]$, the existence condition for symmetry, $ \pounds_{\tilde X_{\bar t}}\tilde L_{\tilde t} = 0 $, leads to the following system of partial differential equations
\bes\label{e5}
\alpha + 2\tilde a \alpha_{,\tilde a}= 0;~~~~3\alpha + 2\tilde a \beta_{,\tilde\phi}= 0;~~~~6\alpha_{,\tilde\phi}-\kappa{\tilde a}^2\beta_{,\tilde a} = 0;\\& 3\alpha \left(\frac{k}{\kappa}-\tilde a^2\tilde V+w\rho_0\tilde a^{-(3w+1)}e^{-{\sqrt {\kappa\over 6}}(1-3w)\tilde \phi}\right)
+\beta\left[\sqrt{\kappa\over 6}(1-3w)\rho_0\tilde a^{-3w}e^{-{\sqrt {\kappa\over 6}}(1-3w)\tilde \phi}-\tilde a^3 \tilde V_{,\tilde\phi}\right]=0, \end{split}\ee
which may be solved to obtain the symmetries listed underneath (see appendix).\\
1. $k = 0, \pm 1$ in vacuum and in radiation era, $F= F_0 R^2$ and $\mathfrak{F}= \beta_0{\tilde a^3} {\dot{\tilde\phi}} = a^3 \dot R$.\\
2. $k = 0$, in vacuum, $F = {F_0\over R}$ with $\mathfrak{F} = \left[\sqrt{1\over \tilde a}\left(-\frac{6\tilde a \dot{\tilde a}}{\kappa}\right)
+\sqrt{6\over\kappa}\tilde a^\frac{3}{2}\dot {\tilde \phi}\right]\exp\left({-\sqrt{3\kappa\over 8}\tilde \phi}\right)= R\dot a\sqrt a$.\\
3. $k = 0$, in vacuum, $F(R) = F_{0}R^{\frac{7}{5}}$, with $\mathfrak{F}= \left[\sqrt{1\over \tilde a}\left(-\frac{6\tilde a \dot{\tilde a}}{\kappa}\right)
-\sqrt{6\over\kappa}\tilde a^\frac{3}{2}\dot {\tilde \phi}\right]\exp\left({\sqrt{3\kappa\over 8}\tilde \phi}\right)=\sqrt a {d\over dt}(a R^{2\over 5})$.\\
In the above, we have translated the  conserved currents to proper time, using the transformation relations \eqref{e1}, which in the RW metric \eqref{rw} under consideration takes the form,
\be\label{e1.1}\tilde a = {\sqrt {F'}} a ={\sqrt {\Phi}} a,\;\;\;\; \;\;\;  d\tilde t = {\sqrt {F'}} dt= {\sqrt {\Phi}} dt.\ee
The most important and expected outcome is, Noether counterpart of the general symmetry \eqref{3t4} has been realized only here, from Einstein's frame action. We have also found some additional symmetries in Einstein's frame, and lost the cherished \cite{b1} one, viz. $F=F_0 R^{3\over 2}$, which carries a conserved current $\mathfrak{F} = {d\over dt}(a\sqrt R)$ both in vacuum and dust era.\\

\subsection{Transforming symmetries back and forth between Jordan's and Einstein's frame}

We have observed that the two frames don't admit same Noether symmetries. Therefore let us check, if the symmetries may be transformed back and forth between the two frames. Using relations \eqref{2rel}, \eqref{e1} and \eqref{e1.1} one finds
\bes\label{4rel}
{\dot{\tilde\phi}}= \sqrt{3\over 2\kappa}\frac{d}{dt}(\ln{\Phi})\frac{dt}{\tilde dt} = \sqrt{3\over 2\kappa}{\Phi}^{-\frac{3}{2}}{\dot\Phi},~~~~
\dot{\tilde{a}}  =\frac{d}{ dt}({\sqrt {\Phi}}a)\frac{dt}{\tilde dt}=\dot{a}+\frac{a\dot{\Phi}}{2\Phi}.\end{split}\ee
Now transforming the conserved current $\mathfrak{F} = \beta_0{\tilde a^3} {\dot{\tilde\phi}}$, for $F(R) = F_{0}R^2$, found in Einstein's frame in vacuum and radiation era, to Jordan's frame, we get $\mathfrak{F}_j = a^3 \dot{\Phi}$. This although satisfies the field equations in Jordan's frame, doesn't satisfy the associated Noether equations \eqref{j8}. The same is true for all other conserved currents obtained in one frame or the other. Thus the symmetries can't be transformed back and forth between the two frames, which clearly indicates that the two are mathematically inequivalent. One may argue that Noether symmetry depends on the choice of configuration space variables. This argument doesn't hold under the present circumstances, since all the results obtained are the associated symmetries of action \eqref{1a}. Therefore if we admit that both the canonical actions \eqref{j3}, and \eqref{e2} truly represent action \eqref{1a}, then both must either independently admit all the symmetries involved with action \eqref{1a}, or the symmetries must be translated from one frame to the other. \\

To understand the reason for such non-congruence, let us now translate Noether equations \eqref{e5} obtained in Einstein's frame to the Jordan's frame, using the transformation relations \eqref{e1} and of-course \eqref{e1.1}. The resulting equations are

\bes\label{e2j}
\alpha + 2 a \alpha_{,a}= 0;~~~~3\alpha + \sqrt{8\kappa\over 3} a{\Phi}^{\frac{1}{2}} \left(\Phi\beta_{,\Phi}-\frac{a\beta_{,a}}{2}\right)= 0;\\&
 6\sqrt{2\kappa\over 3}\left(\Phi \alpha_{,\Phi}-\frac{a\alpha_{,a}}{2}\right)-\kappa{a}^2\left({\Phi}^{\frac{1}{2}}\beta_{,a} +\sqrt{2\kappa\over3}{\Phi}^2\beta_{,\Phi}\right) = 0;\\&
 3\alpha \left( \frac{a^2 U}{2\Phi} - k-\frac{w\kappa\rho_0 a^{-(3w+1)}}{\Phi}\right)
 + \sqrt{\kappa\over 6} a^3{\Phi}^{\frac{1}{2}}\beta \left( U_{,\Phi}- \frac{2 U}{\Phi}-\frac{(1-3w)\kappa\rho_0 a^{-3(w+1)}}{{\Phi}^\frac{3}{2}}\right) =0.
\end{split}\ee

\noindent
The above Noether equation's do not match with those \eqref{j8}, obtained in view of the point Lagrangian \eqref{j6} corresponding to Jordan frame of reference. The same is true for the opposite. Thus, unlike classical field equations, Noether equations can't be translated back and forth, indicating that the two frames are not truly mathematically equivalent. Further, since all the symmetries found in one frame or the other, satisfy the Robertson-Walker counterpart of the field equations \eqref{3f}, corresponding to action \eqref{1a}, hence it is apparent that both the frames only partially represent action \eqref{1a} and therefore have limitations. \\

\section{Quantum Aspect}

Additionally, the two frames represent different quantum dynamics. For example, the phase-space structure of the Hamiltonian in Einstein frame of reference may be expressed as
\bes\label{He}
H_E=-{\kappa\over 12\tilde{a}}P_{\tilde{a}}^2+{1\over 2\tilde{a}^3}P_{\tilde\phi}^2-{3\kappa\over k}\tilde{a}+\tilde{a}^3\tilde V({\tilde{\phi}})
+\rho_0\tilde a^{-3w}e^{-{\sqrt {\kappa\over 6}}(1-3w)\tilde \phi}
\end{split}.\ee
Thus canonical quantization leads ($\hat H \Psi = 0$, due to reparametrization invariance) to

\bes\label{Qe}
\Bigg[\frac{\hbar^2}{2}\left({\kappa\over 6\tilde{a}}\frac{\partial^2}{\partial \tilde{a}^2} -{1\over \tilde{a}^3}\frac{\partial^2}{\partial {\tilde\phi}^2}\right)
+W(\tilde{a},\tilde{\phi})\Bigg]\Psi =0
\end{split}\ee
where,
\be\label{Pe}W(\tilde{a},\tilde{\phi}) =-\left({3\kappa\over k}\tilde{a}-\tilde{a}^3\tilde V({\tilde{\phi}})-\rho_0\tilde a^{-3w}e^{-{\sqrt {\kappa\over 6}}(1-3w)\tilde \phi} \right)\ee
On the contrary, the phase-space structure of the Hamiltonian in Jordan frame of reference, is expressed as,
\bes\label{Hj}
H_J=-{\kappa\over 3{a}^2}P_{{a}} P_\Phi +{\kappa \Phi\over 3{a}^3}P_{\Phi}^2-{3ka{\Phi}\over \kappa}+{{a}^3\over2\kappa} U({\Phi})+\rho_0 a^{-3w}
\end{split}.\ee
Thus canonical quantization leads to

\bes\label{Qj}
\Bigg[{\kappa\hbar^2\over3 a^2}\left({\partial^2\over\partial a\partial \Phi}- {\Phi\over a}\frac{\partial^2}{\partial {\Phi}^2}\right)
+M(a,\Phi)\Bigg]\Psi =0
\end{split}\ee
where,

\be\label{Pj}M(a,\Phi) =-{3ka{\Phi}\over \kappa}+{{a}^3\over2\kappa} U({\Phi})+\rho_0 a^{-3w}. \ee
Operator ordering ambiguities appear in the very first term of \eqref{Qe}, and in the first and second terms of equation \eqref{Qj} which have not been resolved, since, as we shall see later, it is not required at this stage for the present analysis. Now let us transform the quantum equation obtained in Einstein frame \eqref{Qe} to Jordan frame of reference. In view of the transformation relations \eqref{2rel}, \eqref{e1} and \eqref{e1.1}, the inverse transformation relations read

\be a=\tilde a e^{-{1\over 2}\sqrt{2\kappa\over 3}\tilde\phi};\;\;\Phi = e^{\sqrt{2\kappa\over 3}\tilde\phi}.\ee
Therefore,

\be {\partial a\over\partial\tilde a} = {1\over\sqrt\Phi};\;\;{\partial a\over\partial\tilde \phi}=-\sqrt{\kappa\over 6}a;\;\;{\partial \Phi\over\partial\tilde a}=0;\;\;{\partial \Phi\over\partial\tilde \phi} = \sqrt{2\kappa\over 3}\Phi.\ee

\be\label{t1} {\partial\over \partial\tilde a}= \left({\partial a\over \partial\tilde a}\right){\partial\over \partial a}+\left({\partial \Phi\over \partial\tilde a}\right){\partial\over \partial \Phi}={1\over \sqrt\Phi}{\partial\over \partial a};\;\;\;\; {\partial^2\over \partial\tilde a^2}= {1\over\Phi}{\partial^2\over\partial a^2}.\ee
Similarly,
\bes\label{t2}{\partial\over \partial\tilde\phi}= -\sqrt{\kappa\over 6}a{\partial\over \partial a}+\sqrt{2\kappa\over 3}\Phi{\partial\over\partial\Phi}
{\partial^2\over \partial\tilde\phi^2}\\&=
{\kappa\over 6}a^2{\partial^2\over\partial a^2}-{2\kappa\over 3}a\Phi{\partial^2\over\partial a\partial\Phi}+{2\kappa\over 3}\Phi^2{\partial^2\over\partial\Phi^2}
+{2\kappa\over 3}\Phi{\partial\over\partial\Phi}+ {\kappa\over 6}a{\partial\over\partial a}\end{split}\ee
Finally, the potential may be transformed in view of \eqref{j2} and \eqref{e3} as

\bes\label{t3} \tilde V(\tilde\phi)= {R F' - F\over 2\kappa F'^2};\;\;\; U(\Phi) = \chi\Phi - F = R F' - F;\;\;\; \mathrm{So}, \;\;\;
\tilde V(\tilde\phi) = {U(\Phi)\over 2\kappa\Phi^2}.\end{split}\ee
In view of the above transformation relations \eqref{t1} through \eqref{t3}, equation \eqref{Qe} takes the form

\bes\label{Qjt}\left[{\kappa\hbar^2\over a^2}\left({2\over 3}{\partial^2\over\partial a\partial \Phi}-{1\over 6\Phi}{\partial\over\partial a} - {2\over 3 a}{\partial\over\partial\Phi} -{2\Phi\over 3}{\partial^2\over\partial\Phi^2}\right) + M(a,\Phi)\right]\Psi
 = 0\end{split}.\ee
Interestingly, although the potential $W(\tilde a, \tilde \phi)$ transforms correctly from one frame to the other $M(a, \Phi)$, the kinetic part is quite different from \eqref{Qj}. This clearly dictates that something is wrong in connection with the transformation of momenta. The two Hamiltonians \eqref{Qj} and \eqref{Qjt} are quite different, not being related under canonical transformation, and equating through operator ordering clearly doesn't make sense. Therefore, the frames are dynamically inequivalent in the quantum domain. The two frames represent two different systems, as we shall see next.

\section{Possible reason behind in-equivalence}
We have expatiated the fact that the two frames are mathematically inequivalent. In this section we explore the reason behind such in-equivalence. Although the actions are invariant under transformation, translating the point Lagrangian \eqref{e4} obtained in Einstein's frame to the Jordan's frame, using transformation relations \eqref{e1.1}, \eqref{4rel} and \eqref{t3}, we find
\be \tilde L_{\tilde t} = {L_t(a,\Phi, \dot a, \dot\Phi)\over \sqrt \Phi}.\ee
An equivalent relation had also been explored earlier in connection with non-minimally coupled scalar-tensor theory of gravity \cite{18}. Clearly, under transformation, Einstein's frame leads to a point Lagrangian, different from \eqref{j6}. One can also find the canonical momenta in view of the point Lagrangian \eqref{j6} and \eqref{e4} and transform from Einstein's frame to the Jordan (say) to obtain
\bes \label{mom} \tilde {P}_{\tilde a} =  -\frac{6\tilde {a}{{\dot{\tilde a}}}}{\kappa}=-\left[\frac{3a^2{\Phi}^{-\frac{1}{2}}\dot{\Phi}+6a\dot{a}{\Phi}^\frac{1}{2}}{\kappa}\right] = {P_{a}\over\sqrt\Phi}\\&
\tilde{P}_{\tilde\phi}= \tilde{a}^3\dot{\tilde\phi} = \sqrt{3\over2\kappa}{a}^3\dot{\Phi} = \sqrt{2\kappa\over 3}\left(\Phi P_{\Phi} - {a\over 2}P_a\right).\end{split}\ee
Therefore,
\be \label{h} H_E = {H_J\over\sqrt \Phi}.\ee
The, Hamiltonian in Einstein's frame clearly represents a different system altogether, since $H_J$ and ${H_J\over \sqrt \Phi}$ are not related under canonical transformation. In fact the two match only for $\Phi = 1 = F'(R)$, i.e. for General Theory of Relativity and not for higher order theory of gravity. Hence, the two frames are mathematically in-equivalent, and so, physical in-equivalence is obvious. It is important to mention that due to diffeomorphic invariance, in the classical domain the ($^0_0$) equations ($H_E = 0 = {H_J\over \sqrt\Phi} = H_J$) are the same. However, during canonical quantization one can't get rid of $\sqrt\Phi$ appearing in \eqref{h}, since ${H_J\over\sqrt \Phi}$ now acts as the operator, which involves operator ordering. Therefore, at the end the two frames yield completely different quantum descriptions, which has been proved earlier at one loop level calculations \cite{19}.\\

\noindent
Besides the fact that the Lagrangian, the phase-space Hamiltonian and the Noether equations don't match, field equations do, definitely appear interesting. The underlying reason being, gravity is a constrained theory due to the diffeomorphic invariance, which in the present situation (Robertson-Walker model) results in Hamiltonian constraint (Momenta constraints appear, when the metric contains time-space components). In fact it has been proved earlier that Euler-Lagrange equations for conformal Lagrangians transform co-variantly under the conformal transformation relating the Lagrangians, if and only if Hamiltonian vanishes \cite{20}. To understand the situation, let us take the following toy model, represented by the Lagrangian

\be L_1 = {1\over \sqrt x}L;\;\;\;\text{where},\;\;\; L = {m\over 2}\dot x^2 - V(x).\ee
Equation of motion and the Hamiltonian (in terms of configuration space variables) are
\be m{\ddot x} + V'-{m\dot x^2\over 4 x}-{V\over 2x} = 0.\ee
\be H_1 = \frac{m\dot x^2}{2\sqrt x} + {V\over \sqrt x}\ee
are clearly different from those, which appears from $L$. However, if the Hamiltonian would have been constrained to vanish, as in the case of gravity, the equation of motion would remain unaltered, in view of $H_1 = 0$. Further, treating $H = 0$ as the $(^0_0)$ equation of Einstein, the last equation also matches. On the contrary, in terms of the phase-space variables, the Hamiltonian reads,
\be H_1 = {\sqrt x p_x^2\over 2m} + {V\over \sqrt x}\ee
which is clearly different from $H = {p_x^2\over 2m} + V$, and involves operator ordering yet again. Note that despite the fact that the equations of motion corresponding to $L_1$ turned out to be the same as for $L$, it's not possible to make a canonical transformation so that $H_1 = H$, since $H$ and $H_1$ represent two different systems yielding different Hamilton's equations of motion viz.,

\be \label{ne}\dot x = {\sqrt x p_x\over m},\;\;\;\text{and},\;\;\; \dot p_x = -{p_x^2\over 4m\sqrt x} - {V'\over \sqrt x} + {V\over 2 x^{3\over2}}.\ee
rather than ($\dot x = {p_x\over m}$ and $\dot p_x = -V'$). In the above, prime represents derivative with respect to $x$. Further, the Noether equation in this case reads

\be \alpha\left[-{m\dot x^2\over 4 x}+ {V \over 2 x} - {V'}\right] + m \dot x^2 {d\alpha\over dx} =0,\ee
instead of $m \dot x^2 {d\alpha\over dx} - \alpha V' = 0$, leading to $V = V_0$ - a constant, which only assures conservation of momentum. In the present situation \eqref{ne}, the co-efficient of $\dot x^2$ yields $\alpha = \alpha_0 x^{1\over 4}$, and the other equation yields $V = V_0 \sqrt x$. The associated conserved current is therefore, ${\mathcal F} = \alpha_0 m{\dot x\over x^{1\over 4}}$. Time derivative of the conserved current satisfies the equation of motion under the choice $\alpha_0 = 1$, for the form of the potential so obtained. Further, both the equations of motion $m\ddot x + V'= 0$ and ${1\over 2}m \dot x^2 + V = 0$, are also satisfied under the additional condition $V_0 = -{m\over 2}{\mathcal F}^2$. This is exactly the same situation that we have encountered in the Einstein's frame description of $F(R)$ theory of gravity. Under conformal transformation, a different system with a different Lagrangian has evolved. Field equations are also different, but matches due to the presence of the constraint (diffeomorphic invariance). The $(^1_1) = (^2_2) = (^3_3)$ and $(^0_0)$ equations of Einstein, obtained in Einstein's frame match the corresponding ones obtained in Jordan's frame under the transformation relations \eqref{e1.1}, \eqref{4rel} and \eqref{t3}. However, the $\tilde\phi$ equation \eqref{fe2} doesn't match the $\Phi$ equation \eqref{j5}, unless one further uses $(^0_0)$ and $(^1_1)$ equations, as demonstrated below. The field equations corresponding to Einstein's frame are;

\be\label{ee1} 2{\ddot {\tilde a}\over\tilde a} + {\dot{\tilde a}^2\over \tilde a^2} + {k\over \tilde a^2}+\kappa\Big({1\over 2}\dot{\tilde\phi}^2 -\tilde {V}(\tilde \phi)\Big) + \kappa w\rho_0\tilde a^{-3(w+1)}e^{-{\sqrt {\kappa\over 6}}(1-3w)\tilde \phi} = 0.\ee
\be\label{ee2} \ddot{\tilde \phi} + 3 {\dot{\tilde a}\over \tilde a}\dot{\tilde \phi} + {\partial {\tilde V}\over \partial{\tilde \phi}} - \sqrt{\kappa\over 6} (1-3w)\rho_0\tilde a^{-3(w+1)}e^{-{\sqrt {\kappa\over 6}}(1-3w)\tilde \phi} = 0\ee
\be\label{ee3} 3\left({\dot{\tilde a}^2\over \tilde a^2} + {k\over \tilde a^2}\right) - \kappa\Big({1\over2}\dot{\tilde\phi}^2 +\tilde {V}(\tilde \phi)\Big) - \kappa \rho_0\tilde a^{-3(w+1)}e^{-{\sqrt {\kappa\over 6}}(1-3w)\tilde \phi} = 0.\ee
Now, using the transformation relations \eqref{e1.1}, \eqref{4rel} and \eqref{t3}, field equations \eqref{ee1} and \eqref{ee3} transform to,

\be\label{je1} 2{\ddot a\over a} + {\dot{a}^2\over a^2} + {k\over a^2}+ {\ddot \Phi\over \Phi}+ 2{\dot a\dot\Phi\over a\Phi}- {U(\Phi)\over 2\Phi} + \kappa w\rho_0{a^{-3(w+1)}\over\Phi} = 0,\ee

\be\label{je3} 3\left({\dot a^2\over a^2} + {k\over a^2}\right) + 3 {\dot a\dot\Phi\over a\Phi} - {U(\Phi)\over 2\Phi} - \kappa \rho_0 {a^{-3(w+1)}\over \Phi} = 0,\ee
which are exactly the $(^1_1) = (^2_2) = (^3_3)$ and the $(^0_0)$ equations respectively, as may be realized in Jordan's frame of reference. However, under transformation \eqref{ee2} takes the form
\be\label{jne5} {\ddot\Phi\over \Phi} + 3{\dot a\dot \Phi\over a\Phi} + {1\over 3}U_{,\Phi} - {2\over 3}{U(\Phi) \over \Phi} -{\kappa\over 3} (1-3w)\rho_0 {a^{-3(w+1)}\over \Phi} = 0\ee
which is not the $U_{,\Phi} = R$ equation obtained in \eqref{j5}. This clearly dictates that the field equations are also not the same under conformal transformation. This fact had also been noticed earlier in connection with non-minimally coupled scalar-tensor theory of gravity \cite{18}. Nevertheless as already mentioned, the diffeomorphic invariance of gravity, which is essentially the $(^0_0)$ equation of Einstein, saves the soul. Using, \eqref{ee1} and \eqref{ee3}, one realizes \eqref{j5} from equation \eqref{jne5}. Note that unlike the toy model, here one requires to use $(^1_1)$ equation in addition due to the presence of coupling, as already proved earlier \cite{18, 20}. Noether equations are obviously different, and yield new conserved quantities, which satisfy the field equations as well under some constraints relating the constants appearing in the potential with the coupling parameter.

\section{Concluding Remarks}

Whether every non-minimally coupled scalar-tensor theory of gravity is conformally equivalent to minimally coupled one, is a long standing debate. In connection with non-minimally coupled scalar-tensor theory of gravity, it had been noticed earlier that conformal transformation connecting Jordan's and Einstein's frames does not preserve Noether symmetry ($\pounds_{X_t}L_t \ne \pounds_{\tilde X_{\tilde t}}\tilde L_{\tilde t}$), when the time coordinate is the cosmic time, since $d\tilde t = \sqrt{-2f(\phi)}~ dt$, where $f(\phi)$ is the coupling parameter \cite{21}. The authors \cite{18, 21} also established the fact that Noether symmetry remains preserved if conformal time $\eta$ is chosen as the time coordinate instead, where $a(\eta)^2 d\eta^2 = dt^2$, i.e. $\pounds_{X_\eta}L_\eta = \pounds_{\tilde X_\eta}\tilde L_\eta$. However, firstly, the equivalence is broken in the presence of ordinary matter, and second, to analyse the phenomenology relative to a given model and to then obtain quantities comparable with observational data, the appropriate time coordinate is the cosmic time. Nevertheless, the authors \cite{18, 21} further established the fact that Noether symmetry is preserved with cosmic time under generalized form $\pounds_{X_t}L_t - {\pounds_{X_t} a\over a} E_t = \pounds_{\tilde X_{\tilde t}}\tilde L_{\tilde t} - {\pounds_{\tilde X_{\tilde t}} \tilde a\over \tilde a} \tilde E_{\tilde t}$, where $E_t$ and $\tilde E_{\tilde t}$ are the energy functions in Jordan's and Einstein's frames respectively. However, this is too intuitive and holds for homogeneous and isotropic model only, and also the presence of ordinary matter field breaks the equivalence, as mentioned earlier.\\

Here, in connection with the symmetry analysis for higher order theory of gravity, we have clearly explored the fact that, the Jordan's frame and Einstein's frame of references are not mathematically equivalent, unless $\Phi = F'(R) = 1$, which is essentially General theory of relativity. In view of earlier discussion, it is clear that, conformal time must not be invoked to establish equivalence. The field equations match, since the Hamiltonian is constrained to vanish. Further, Noether equations and the symmetries don't match, since, as we understand, Noether equations don't administer the constraint of a theory. Hence the two frames differ mathematically and hence dynamically.\\

In order to unify early inflation with late-time acceleration in view of $F(R)$ theory of gravity, Einstein's frame is invoked at the early stage of cosmic evolution, since the resulting inflationary parameters (the spectral index of density perturbation $n_s$, and the scalar to tensor ratio $r$) are consistent with experimental (WMAP-Bao-SN1a-Planck) data. Further, to pass the cosmological bound and the solar test simultaneously, Chameleon mechanism is invoked and the mass of the so-called scalaron field is inspected, for which again Einstein's frame of reference is recalled \cite{a, b}. This technique is therefore under serious challenge, and so $F(R)$ theory of gravity might not be treated as an alternative to dark energy.\\

Nevertheless, if one can find a symmetry generator which somehow involves the Hamiltonian constraint, then it might be possible to establish equivalence between the two frames. This we pose in future. Even if it's possible to establish classical equivalence between the two frames, the phase-space structure of the two Hamiltonian are different, leading to a totally different quantum description of the theory.\\

At the end, we would like to mention that, being a higher order theory, $F(R)$ requires additional degree of freedom for canonical formulation. However, canonical quantization requires to choose appropriate basic variables viz., $h_{ij}$ and $K_{ij}$, which are the induced three space metric and the extrinsic curvature tensor respectively. The scalar-tensor equivalent forms on the contrary choose $\Phi$ or $\tilde\phi$ in addition to $h_{ij}$. In the process, the Ricci scalar ($R$) is treated as an additional basic variable, which can't be translated to $K_{ij}$ following canonical transformation. Therefore, not being a physical variable, such quantization schemes using the Ricci scalar ($R$) are not viable.\\

\section{Appendix}

In the appendix, we produce detailed calculation regarding the symmetries obtained in Jordan's and Einstein's frames of references.

\subsection{Noether symmetry in Jordan's Frame}
The first three of the set of equations \eqref{j8} may be solved to obtain
\be\label{jns} \alpha = {\alpha_1\over a}+ \alpha_2 a; ~~~~~~~\beta = {\beta_0\over a} - \left({\alpha_1\over a^2}+ 3 \alpha_2\right) \Phi,\ee
where, $\alpha_1~\alpha_2$ and $\beta_0$ are constants of integration. However, the above forms of $\alpha$ and $\beta$ don't satisfy the last of the set \eqref{j8}, in general. It is interesting to note that the set of Noether equations are the same in both the vacuum era ($\rho_0 = 0$), and in pressure-less dust era ($p = 0 = w$). Therefore, in both the era, same solution is admissible. Now, only under the choice $\alpha_2 = 0 = \beta_0$, the set of equations (\ref{j8}) may be solved for arbitrary curvature parameter $k = 0, \pm 1$, in both the vacuum and dust era as,
\be\label{2.10}
\alpha={\alpha_{1}\over a},~~~ \beta=-\frac{\alpha_{1}\Phi}{a^2},~~~ \ U = U_0\Phi^3.
\ee
In view of the transformation relations (\ref{j2}), one therefore obtains,

\be\label{2.11} F(R) = F_{0} R^{\frac{3}{2}},\ee
and the corresponding conserved current reads

\be\label{2.21} \mathfrak{F} = [a \dot\Phi + \dot a \Phi]= \frac{d}{dt}(a\Phi) =\frac{d}{dt}(a\sqrt R).\ee
In the radiation dominated era, however ($\rho = 3p=\rho_0a^{-4}$), the set of Noether equations (\ref{j8}) does not admit any solution.

\subsection{Noether symmetry in Einstein's Frame}
The first three of the set of equations \eqref{e5} may be solved to obtain
\bes\label{ens}
    \alpha = \sqrt{1\over \tilde a} \left[A\exp\left({\sqrt{3\kappa\over 8}\tilde \phi}\right)+ B \exp\left(-{\sqrt{3\kappa\over 8}\tilde \phi}\right)\right] \\&
    \beta = - \sqrt{6\over\kappa \tilde a^3} \left[A\exp\left({\sqrt{3\kappa\over 8}\tilde \phi}\right) - B \exp\left(-{\sqrt{3\kappa\over 8}\tilde \phi}\right)
     \right]+\beta_0,
\end{split}\ee
where, $A,~B$ and $\beta_0$ are constants of integration. However, the above forms of $\alpha$ and $\beta$ don't satisfy the last of the set \eqref{e5}, in general. Nevertheless, some special forms of $\alpha$ and $\beta$ satisfy the last equation in different epoch of cosmic evolution, which we explore in the following subsections.

\subsubsection{Noether symmetry in pure vacuum}

The last of the set of equations \eqref{e5} in vacuum ($\rho_0 = 0$) reads
\be \label{el} 3\alpha \left(\frac{k}{\kappa}-\tilde a^2\tilde V\right)-\beta\tilde a^3 \tilde V_{,\tilde\phi}=0.
\ee
\textbf{Case-I ($k = 0, \pm 1$)}\\

\noindent
Now under the choice $A = 0 = B$, \eqref{ens} and \eqref{el} yield,
\be\label{e10}
    \alpha = 0,~~~~~\beta = \beta_0~~~~~ \tilde V(\tilde\phi) = V_{0},\ee
while the conserved current reads,
\be\label{e11}\mathfrak{F} = \beta_0{\tilde a^3} {\dot{\tilde\phi}}.\ee
In view of \eqref{e3}, the form of $F(R)$ is therefore obtained as,
\be\label{e12} F(R) = F_{0}R^2, \ee
under the condition, $8\kappa V_0 F_0 = 1$. Transforming the above conserved current \eqref{e11} in proper time, using the relation between $\tilde\phi$ and $F'(R)$ given in \eqref{e1}, the transformation relations \eqref{e1.1}, and the above form of $F(R)$ (\ref{e12}), one obtains,
\be\label{e14} a^3 \dot R = \mathfrak{F}.\ee

\noindent
\textbf{Case-II ($ k = 0$)}\\

\noindent
Choosing $\beta_0 = 0$, the expression of the potential may be found from \eqref{el} as,
\bes\label{e32}
 \tilde V(\tilde\phi) =  V_{0}\left[A\exp\left({\sqrt{3\kappa\over 8}\tilde \phi}\right)- B \exp\left(-{\sqrt{3\kappa\over 8}\tilde \phi}\right)\right]^2.
     \end{split}\ee
Unfortunately, the above form of the potential \eqref{e32} does not yield any form of $F(R)$ in general. We therefore study some special cases.\\

\noindent
\textbf{Case-II.a}\\

\noindent
Setting the constant $A = 0$, the solutions \eqref{ens} and \eqref{e32} read
\be\label{e18}
 \alpha = B\sqrt{1\over \tilde a} \exp\left({-\sqrt{3\kappa\over 8}\tilde \phi}\right),~~~~~\beta = B\sqrt{6\over\kappa \tilde a^3} \exp\left({-\sqrt{3\kappa\over 8}\tilde \phi}\right),~~~~~\tilde V(\tilde\phi) =  V_{0}B^2\exp\left({-\sqrt{3\kappa\over 2}\tilde \phi}\right)\ee
and the expression for the associated conserved current is
\be\label{e19} \mathfrak{F} = \left[\sqrt{1\over \tilde a}\left(-\frac{6\tilde a \dot{\tilde a}}{\kappa}\right)
+\sqrt{6\over\kappa}\tilde a^\frac{3}{2}\dot {\tilde \phi}\right]\exp\left({-\sqrt{3\kappa\over 8}\tilde \phi}\right)\ee
Now in view of relations \eqref{e1}, \eqref{e3} and the Noether potential \eqref{e18} the form of $F(R)$ is found as
\be\label{e20} F(R) =  \frac{ F_{0}}{R}. \ee
Finally, using the transformation relation \eqref{e1}, \eqref{e1.1} and the above form of $F(R)$, as before, the conserved current \eqref{e19} may be translated in proper time as
\be\label{e21}  R\sqrt{a}~ \dot a =\mathfrak{F} \ee
Following a little algebraic computation, one can check that the conserved current \eqref{e21} so obtained satisfies the field equations corresponding to the form of $F(R)$ \eqref{e20}, which are,

\bes\label{e23}2\frac{\ddot a}{a} + \frac{\dot a^2}{a^2} - 4 \frac{\dot a\dot R}{a R}-2\frac{\ddot R}{R}+6\frac{\dot R^2}{R^2}- R=0,~~~~~~3\frac{\dot a^2}{a^2} - 6\frac{\dot a\dot R}{a R}- R =0.\end{split}\ee
under the condition, $F_0 = -\kappa^2 V_0^2 B^4$. It is also interesting to note that the solution for the scale factor may also be obtained immediately in view of the above conserved current as, $a(t)= a_0 t^2$. Thus $1\over R$, term which was put in by hand as an alternative to the dark energy, is an outcome of Noether symmetry of higher order theory of gravity. However, such form is admissible in pure vacuum, and therefore appears to be suitable for power law inflation, rather than late time acceleration.\\

\noindent
\textbf{Case-II.b}\\

\noindent
Setting $B = 0$ on the other hand, one obtains
\be\label{e25}
 \alpha = A\sqrt{1\over \tilde a} \exp\left({\sqrt{3\kappa\over 8}\tilde \phi}\right),~~~~~\beta = - A\sqrt{6\over\kappa \tilde a^3} \exp\left({\sqrt{3\kappa\over 8}\tilde \phi}\right),~~~~~\tilde V(\tilde\phi) =  V_{0}A^2\exp\left({\sqrt{3\kappa\over 2}\tilde \phi}\right)\ee
and the expression of Conserved current reads,
\be\label{e26} \mathfrak{F} = \left[\sqrt{1\over \tilde a}\left(-\frac{6\tilde a \dot{\tilde a}}{\kappa}\right)
-\sqrt{6\over\kappa}\tilde a^\frac{3}{2}\dot {\tilde \phi}\right]\exp\left({\sqrt{3\kappa\over 8}\tilde \phi}\right)\ee
One can find the form of $F(R)$, in view of equations \eqref{el}, \eqref{e3}, \eqref{e25} as,
\be\label{e27} F(R) =  F_{0}R^{\frac{7}{5}} \ee
Finally, using the transformation relation \eqref{e1}, \eqref{e1.1} and the above form of $F(R)$, one can translate the conserved current in proper time as
\be\label{e28} \mathfrak{F} = \left[ a^{\frac{1}{2}} \dot a  R^{\frac{2}{5}}+ \frac{2}{5}a^{\frac{3}{2}} R^{{-}\frac{3}{5}} \dot R\right].\ee
The field equations for the form of $F(R)$ so obtained are
\be\label{e30}2\frac{\ddot a}{a} + \frac{\dot a^2}{a^2} + \frac{4\dot a\dot R}{5a R}+\frac{2\ddot R}{5R}-\frac{6\dot R^2}{25R^2}- \frac {R}{7}=0,~~~~~
 3\frac{\dot a^2}{a^2} + \frac{6\dot a\dot R}{5a R}-\frac {R}{7} =0\ee
It may be checked through a little algebra that the conserved current \eqref{e28} satisfies the above field equations under the condition, $7^7 \kappa^2 F_0^5 V_0^2 A^4  = 5^5$. One can also solve the above conservation current for the scale factor as $a(t)= d_0 t^{\frac{6}{5}}$. Thus, $F(R) \propto R^{\frac{7}{5}}$ is yet another form that has been explored following Noether symmetry in vacuum dominated era. However, such a form, being found in vacuum dominated era, is again suitable for early inflation, rather than late-time cosmic acceleration.

\subsubsection{Noether symmetry of $F(R)$ in radiation era.}

In the radiation dominated era ($p = {1\over 3}\rho$), The last of the set of equations \eqref{e5} reads,
\be \label{er} 3\alpha \left(\frac{k}{\kappa}-\tilde a^2\tilde V +{\rho_0\over 3}\tilde a^{-2}\right)-\beta\tilde a^3 \tilde V_{,\tilde\phi}=0.
\ee
Under the choice, $A = 0 = B$, one obtains
\be\label{e36}
\alpha = 0,~~~~~~\beta = \beta_0~~~~~~\tilde V(\tilde\phi) = V_0
\ee
and therefore, equations \eqref{e3} and \eqref{e36} give the form of $F(R)$ as
\be\label{e37} F(R) = F_{0}R^2, \ee
under the condition, $8\kappa V_0 F_0 =1$. The expression of conserved current is
\be\label{e38} \mathfrak{F} = \beta_0{\tilde a^3} {\dot{\tilde\phi}} =  a^3 \dot R.\ee
The last equality appears after translating the conserved current in proper time, which satisfies the field equations as well. Thus Noether symmetry analysis in Einstein's frame admits $F(R) = F_{0}R^2$, both in vacuum and radiation era, i.e. for trace-less energy-momentum tensor. Noether symmetry of $F(R)$ with pressure-less dust does not exist.

\end{document}